\documentclass[journal,twoside,web]{ieeecolor}
\usepackage{cite}
\usepackage{amsmath,amssymb,amsfonts}
\usepackage{algorithmic}
\usepackage{graphicx}
\usepackage{textcomp}
\usepackage{url}
\usepackage{amssymb}
\usepackage{color}
\usepackage{makecell}
\def\logoname{LOGO-tmi-web}
\definecolor{subsectioncolor}{rgb}{0,0.541,0.855}
\def\journalname{IEEE Transactions on Medical Imaging}

\def\BibTeX{{\rm B\kern-.05em{\sc i\kern-.025em b}\kern-.08em
    T\kern-.1667em\lower.7ex\hbox{E}\kern-.125emX}}
\begin{document}
\title{Bridging Synthetic and Real Images: a Transferable and Multiple Consistency aided Fundus Image Enhancement Framework}
\author{Erjian Guo, Huazhu Fu, \IEEEmembership{Senior Member, IEEE}, Luping Zhou, \IEEEmembership{Senior Member, IEEE},  \\
and Dong Xu, \IEEEmembership{Fellow, IEEE}
\thanks{This work is partially supported by the Huazhu Fu's Agency for Science, Technology and Research (A*STAR) Central Research Fund, and Advanced Manufacturing and Engineering (AME) Programmatic Fund (A20H4b0141). L.~Zhou is supported by DP200103223 funded by Australian Research Council (ARC). This work was supported by the funding from The Hong Kong Jockey Club Charities Trust (No.2022-0174), and the startup funding from The University of Hong Kong.} 
\thanks{E.~Guo and L.~Zhou are with the School of Electrical and Information Engineering, University of Sydney, Australia (e-mail: eguo9622@uni.sydney.edu.au; luping.zhou@sydney.edu.au). }
\thanks{D.~Xu is with the Department of Computer Science, The University of Hong Kong, Pokfulam, Hong Kong (e-mail: dongxu@hku.hk). }
\thanks{H.~Fu is with the Institute of High Performance Computing (IHPC), Agency for Science, Technology and Research (A*STAR), Singapore 138632 (E-mail: hzfu@ieee.org).}}
\maketitle

\begin{abstract}
Deep learning based image enhancement models have largely improved the readability of fundus images in order to decrease the uncertainty of clinical observations and the risk of misdiagnosis. However, due to the difficulty of acquiring paired real fundus images at different qualities, most existing methods have to adopt synthetic image pairs as training data. The domain shift between the synthetic and the real images inevitably hinders the generalization of such models on clinical data. In this work, we propose an end-to-end optimized teacher-student framework to simultaneously conduct image enhancement and domain adaptation. The student network uses synthetic pairs for supervised enhancement, and regularizes the enhancement model to reduce domain-shift by enforcing teacher-student prediction consistency on the real fundus images without relying on enhanced ground-truth. Moreover, we also propose a novel multi-stage multi-attention guided enhancement network (MAGE-Net) as the backbones of our teacher and student network. Our MAGE-Net utilizes multi-stage enhancement module and retinal structure preservation module to progressively integrate the multi-scale features and simultaneously preserve the retinal structures for better fundus image quality enhancement. Comprehensive experiments on both real and synthetic datasets demonstrate that our framework outperforms the baseline approaches. Moreover, our method also benefits the downstream clinical tasks.
\end{abstract}


\section{Introduction}
\label{sec:introduction}
\IEEEPARstart{R}{etinal} images are widely used by ophthalmologists or automated image analyzing systems as a non-invasive way to detect and monitor various eye and body diseases~\cite{abramoff2010retinal}, such as glaucoma, diabetic retinopathy, and hypertension. Unfortunately, a study of 5,575 patients found that about 12\% of fundus images are not of adequate quality to be readable by ophthalmologists~\cite{philip2005impact}. The quality of fundus images varies due to equipment limitations, ophthalmologists' experience, and patient eye movement, which could negatively affect clinical decision making. Image enhancement methods are therefore proposed as a remedy. Traditional fundus image enhancement methods~\cite{foracchia2005luminosity,cheng2018structure,setiawan2013color,liao2014retinal} were mainly based on hand-crafted priors, and they could not satisfactorily handle the complexity of varied low-quality cases. To solve this issue, the deep learning methods were proposed to learn more general priors from large amounts of paired low-quality and high-quality images~\cite{shen2020modeling,perez2020conditional,sevik2014identification,LI2021101971,deng2022rformer,Li2022_TMI,Liu2022_MICCAIa,Liu2022_MICCAIb}. Therefore, the existing methods resort to either i) synthetic image pairs, such as synthesizing low-quality fundus images by degrading real high-quality ones~\cite{shen2020modeling}, or ii) unpaired supervision models, such as CycleGAN-like ones~\cite{ma2021structure,deng2022rformer}, for enhancement. However, both approaches have limitations. On one hand, due to the domain shift between the synthetic and the real fundus images, the models trained on synthetic image pairs have limited capability to generalize well to real clinical fundus images. On the other hand, the models trained with unpaired supervision mainly translate image styles and could not well preserve the local details of structures.  

To bridge this gap, in this work, we propose a new end-to-end optimized method that simultaneously conducts image enhancement and domain adaptation in one-shot based on the well-known mean teacher framework~\cite{tarvainen2017mean}. By imitating self-supervised learning, mean teacher framework was proposed to be used for unsupervised domain adaptation task in~\cite{french2017self}.
%French et al.~\cite{french2017self} directly simulated semi-supervised learning to solve unsupervised domain adaptation task. The main idea is to utilize mean teacher framework~\cite{tarvainen2017mean} to work in cross-domain tasks by calculating the consistency loss of two predictions under perturbations of inputs (e.g., different augmentations of image).
The domain gap is naturally reduced by the consistency regularization in the mean teacher framework, which enforces the predictions of the teacher network and the student network to be consistent around each unlabeled (target domain) image. Mean teacher aims to learn a smoother domain-invariant function from unlabeled (target domain) images than the model purely trained on labeled (source domain) images. In this paper, we adapt the mean teacher framework to our cross-domain enhancement network through both multi-stage enhancement consistency and multi-level segmentation consistency. Specifically, our method consists of a student network and a teacher network with identical architecture, while the latter is an exponential moving average of the former. The student network is trained for two tasks. On one hand, it uses synthetic image pairs for supervised enhancement. On the other hand, it uses the unlabeled real images (without the enhanced ground-truth) to regularize the enhancement model trained on the synthetic input, in order to reduce the domain shift. This is achieved by feeding a real image and its augment simultaneously into the student and the teacher networks, respectively, and enforcing consistent predictions between the two networks. 
Moreover, we also propose a powerful multi-stage Multi-Attention Guided Enhancement Network called MAGE-Net, which also serves as the backbones of the student and the teacher network. Our MAGE-Net is comprised of a multi-stage enhancement (MSE) module and a retina structure preservation (RSP) module. The MSE module consists of a UNet-Shaped stage (Stage-1) to encode broad semantic information and an original-scale stage (Stage-2) to provide spatial details. Multi-type attentions are further employed to guide the enhancement, including our newly proposed fundus attention. Compared with the commonly used skipped connections that directly link encoder-decoder levels, our multi-stage multi-attention architecture provides a more delicate way to effectively integrate multi-scale features. The RSP module is proposed to maintain the vital structures information in fundus images, e.g., the vessels, the optic disc, and the cup, for clinical observation. It sequentially produces essential structure features to guide the enhancement process. 
%These structure information at varied resolution levels are merged with the feature maps in MSE module to correct and enhance the useful clinical information. 
Building upon MAGE-Net and the supervised enhancement loss, we further propose multiple consistency losses to bridge the student and teacher networks, including the multi-stage enhancement consistency and the multi-level segmentation consistency of the RSP module.

%Building upon a multi-stage backbone network~\cite{zamir2021multi}, 
%Our backbone enhancement model seamlessly incorporates a simple and effective multi-scale retinal structure preservation submodule. Based on our new network structure, we also design multi-stage enhancement consistency loss and multi-level segmentation consistency losses for the teacher-student framework to reduce the domain gap between the synthetic and the real low-quality fundus images. 
The contributions of this work are summarized as follows:
\begin{enumerate}
    \item We propose a new teacher-student based framework with specifically designed consistency losses to reduce the domain shift between the synthetic and the real low-quality fundus images, which is conducted simultaneously with the image enhancement task.
    \item We propose a new multi-stage multi-attention guided fundus image enhancement network, which corrects low-quality fundus images while catering for contextual accuracy, spatial accuracy, and anatomical structure accuracy. 
    \item The experimental results show that our fundus enhancement method also improves the performance of multiple downstream tasks, such as vessel segmentation, optic disc, and cup detection, and disease recognition.
\end{enumerate}
 
%==========================+%
\section{Related Work}
In this section, we briefly discuss the fundus image enhancement approaches, domain adaptation methods, and the mean teacher framework, which are related to our work.
%-----------------------------------%

\subsection{Fundus Image Enhancement}
 Fundus image enhancement methods have two main categories: prior-based methods and learning-based methods.
\subsubsection{Prior-based methods.}  
 The traditional prior-based methods could not successfully address multiple low-quality cases including noise, blurring, missed focus, illumination, and contrast. Histogram equalization (HE)~\cite{mitra2018enhancement,hsu2015medical,joshi2008colour} is a popular method in this category to improve image contrast of retinal images, but the decreasing of gray levels results in the loss of image details. Therefore, negative observations were found for it in many retina image cases, especially in color retinal images. Alternatively, contrast limited adaptive histogram equalization (CLAHE) is also widely adopted to enhance medical images, e.g., Setiawan et al.~\cite{setiawan2013color} applied a specifically designed CLAHE to retinal fundus images. However, the CLAHE method may produce artificial boundaries at the region containing an abrupt change in the gray levels. Moreover, although these methods perform efficiently due to their simplicity, their heavy dependence on global image statistics leads to severe image degradation in practice. These hand-crafted priors from human observations do not always work in diverse real-world low-quality retinal images. They tend to suffer from undesirable color and structure distortions. 
 
\subsubsection{Learning-based Methods.}
Recently, due to the advantage in image representation, deep learning methods have dominated the computer vision field. There are different types of methods for promoting image quality, such as image enhancement~\cite{ren2019low}, dehazing~\cite{cai2016dehazenet},  denoizing~\cite{huang2021neighbor2neighbor}, and so on. Unfortunately, due to the differences between medical and natural images, the above image correction methods are not suitable to fundus image enhancement which needs specific design to cater for the special characteristic of retinal images. Retinal image enhancement should use pixel-wise translation to preserve retinal structures, which is critical in retinal image analysis. The deep learning networks applied to retinal image enhancement are composed of two categories: synthetic image-pairs-based methods and unpaired-supervision-based methods. The first ones like~\cite{cheng2021secret} require high-low quality retinal image pairs to learn a mapping from one representation to another. The widely used fundus degradation method~\cite{shen2020modeling} simulates real images of low quality to build retinal image pairs from real high-quality images. However, the image pair-based methods ignore the domain gap between the synthetic low-quality images and the real low-quality images, thus generalizing unsatisfactorily to clinic use. The unpaired supervision-based methods~\cite{ma2021structure,shen2020modeling} are usually based on CycleGAN-like frameworks to restore fundus images directly from real unpaired images of high or low quality. However, these methods mainly translate image styles to simulate clean results without well preserving the important details of fundus structures. To this end, vessel segmentation was employed as a useful way to enhance retinal structures. For example, CofeNet~\cite{shen2020modeling}, a method using synthetic image pairs, was designed to preserve the retinal structures in fundus enhancement process through benefiting from vessel segmentation outputs. 
Recently, Transformer-based methods have achieved great success in high-level vision tasks~\cite{Shamshad2022}, such as image classification~\cite{arnab2021vivit}, semantic segmentation~\cite{cao2021swin}, object
detection~\cite{carion2020end}, etc. Due to the advantage of capturing long-range dependencies and good performance in many high-level vision tasks, Transformer has also been introduced into low-level vision tasks, such as image restoration~\cite{liang2021swinir,zamir2022restormer}. The transformer-based method is rarely applied in the fundus image enhancement task. The transformer-based method: RFormer~\cite{deng2022rformer} relies on an in-house Real Fundus (RF) dataset including 120 paired high- and low-quality real fundus images to learn to synthesize high-quality images from low-quality ones. Unlike other fundus image enhancement methods using synthetic low-quality images for training, RFormer is directly trained based on paired real fundus images of different qualities, unfortunately, are costly to be collected in practice. Moreover, the RF dataset has not been publicly released.

In this paper, to alleviate the issues from both the paired- and the unpaired-supervision-based methods and integrate their advantages, we develop a fundus image enhancement framework that learns feature presentations from synthetic image pairs and leverages real low-quality images to improve enhancement performance. This further calls for domain adaptation to alleviate the discrepancy between the synthetic and the real image domains involved in the fundus image enhancement task.
%---------------------------------%
\subsection{Domain Adaptation}
Due to the gap between the source and the target domains, a model trained on the source domain may suffer significant performance drops on the target domain in practice. Domain adaptation is therefore proposed to bridge the domain gap so that the model learned from the source domain is able to perform decently on the target domain. There is a large corporation of literature tackling the problem of domain adaptation. We focus on deep learning based methods as these are most relevant to our work.
Unsupervised domain adaptation could be addressed from different perspectives. Discrepancy-based methods guide the feature learning by minimizing the domain gap with Maximum Mean Discrepancy~\cite{long2017deep}, while the works in~\cite{ganin2015unsupervised,tzeng2017adversarial} estimate the domain confusion by learning a domain discriminator. Differently, self-ensembling~\cite{french2017self} extended the mean teacher framework~\cite{tarvainen2017mean} to reduce domain gap and established several cross-domain benchmarks for recognition task. Recently, Mean Teacher has been extensively used as a transfer learning method for various tasks, e.g., image dehazing~\cite{liu2021synthetic}, object detection~\cite{raj2015subspace} and semantic segmentation~\cite{zhang2018fully}. For example, to reduce the domain gap in image dehazing, Liu et al.~\cite{liu2021synthetic} developed a disentangle-consistency mean-teacher network (DMT-Net) collaborating with unlabeled real-world hazy images to address the domain shift problem. Similar to~\cite{liu2021synthetic}, our method aims to
leverage additional real low-quality fundus images without ground-truth to alleviate the domain discrepancy between the synthetic and the real images. Along this line, we explore the Mean Teacher framework to bridge the domain gap by imposing consistency regularization in fundus image enhancement, which has not been previously explored in this field. Moreover, the enhancement loss and the segmentation loss between synthetic to real fundus image is elegantly integrated into the Mean Teacher paradigm to boost cross-domain enhancement results.

%-----------------------%
\subsection{Mean Teacher Framework}
Mean Teacher framework~\cite{tarvainen2017mean} is widely used in semi-supervised learning.
The main idea of mean teacher is to enforce the predictions of the teacher and the student networks consistent under small perturbations of the input or the network parameters. Mean teacher consists of two networks with the same architecture: let $S(\cdot)$ and $T(\cdot)$ represent the embedding functions of the student network with weight $w_{s}$ and the teacher network with weight $w_{t}$, respectively. Let us denote a labeled data as $\mathbf{I}_l$, an unlabeled data as $\mathbf{I}_u$ and its augment as $\tilde{\mathbf{I}}_u$. The consistency loss penalizes the difference between the student’s prediction $S(\mathbf{I}_u)$ and the teacher’s prediction $T(\tilde{\mathbf{I}}_u)$, which is typically computed as the Mean Squared Error:
\begin{equation}
\mathcal{L}_{cons}(\mathbf{I}_u)=\Vert{S}(\mathbf{I}_u;w_{s}) - {T}(\tilde{\mathbf{I}}_u;w_{t})\Vert_{2}^{2}.
\end{equation}
The student network is trained by using gradient descent, and the
weights $w_{t}$ of the teacher network at the $n$-th iteration are the exponential moving average of the student weights $w_{s}$:
\begin{equation}
w_{t}^{n}=\alpha\cdot w_{t}^{n-1}+(1-\alpha)\cdot w_{s}^{n-1},
\end{equation}
where $\alpha$ is a smoothing coefficient parameter that controls the updating of the teacher's weights. The total loss of the mean teacher framework is a combination of the supervised losses on labeled data and the consistency losses on unlabeled data, balanced with the trade-off parameter $\mu$:
\begin{equation}
\mathcal{L}=\sum^{M}_{i=1}\mathcal{L}_{super}(\mathbf{I}_l^i) +\mu \sum^{N}_{j=1}\mathcal{L}_{cons}(\mathbf{I}_u^j),
\end{equation}
where $M$ and $N$ denote the total number of the labeled and unlabeled images, respectively.

%================================%
\section{Methodology}
\begin{figure*}[!t]
\begin{center}
\centerline{\includegraphics[width=0.9\textwidth]{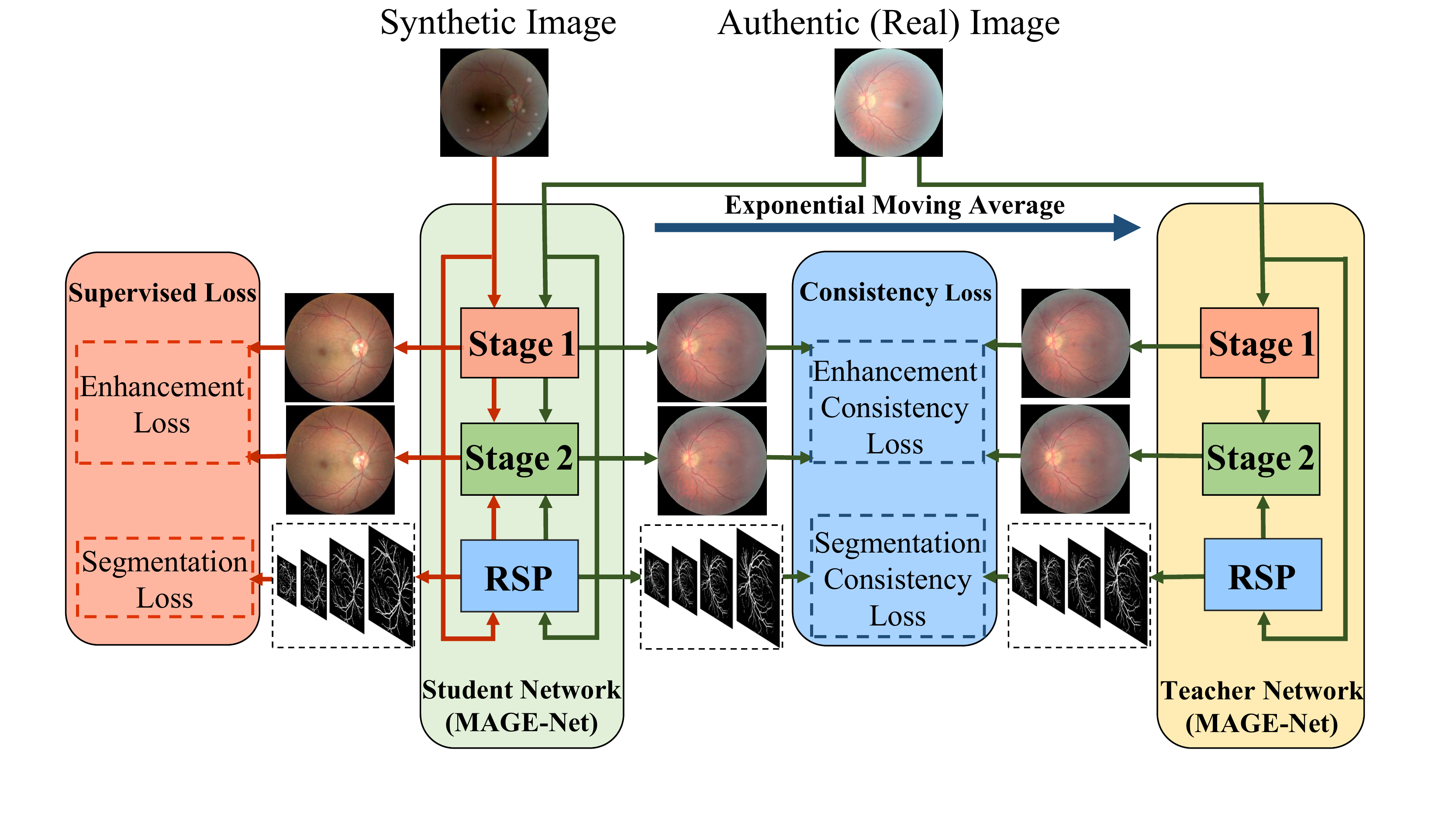}}
\end{center}
\caption{\textbf{Overview of Transferred MAGE-Net with Multi-Stage Consistency (T-MAGE-Net).} We use MAGE-Net as a teacher network and a student network, respectively. Each paired synthetic image is fed into student model to conduct the supervised learning of image enhancement and segmentation. We calculate supervised multi-stage enhancement losses and supervised segmentation losses in the student network. Each unlabeled real image is firstly transformed into two perturbed samples by adding Gaussian noise and then we inject the two perturbed samples into student and teacher models separately. Two types consistency regularization are devised to facilitate reducing the domain gap in mean teacher paradigm: 1) Multi-Stage Enhancement Consistency Loss to align the clean predictions between teacher and student; 2) Multi-Stage Segmentation consistency Loss for matching the retinal structures between teacher and student. The whole T-MAGE-Net is trained by minimizing the supervised losses on paired synthetic image data plus the two consistency losses on the unlabeled real image in an end-to-end manner. Note that the student network is optimized with Adam and the weights of teacher network are the exponential moving average of student model weights.}
\label{fig1}
\end{figure*}

\begin{figure*}[!t]
\begin{center}
\includegraphics[width=0.9\textwidth]{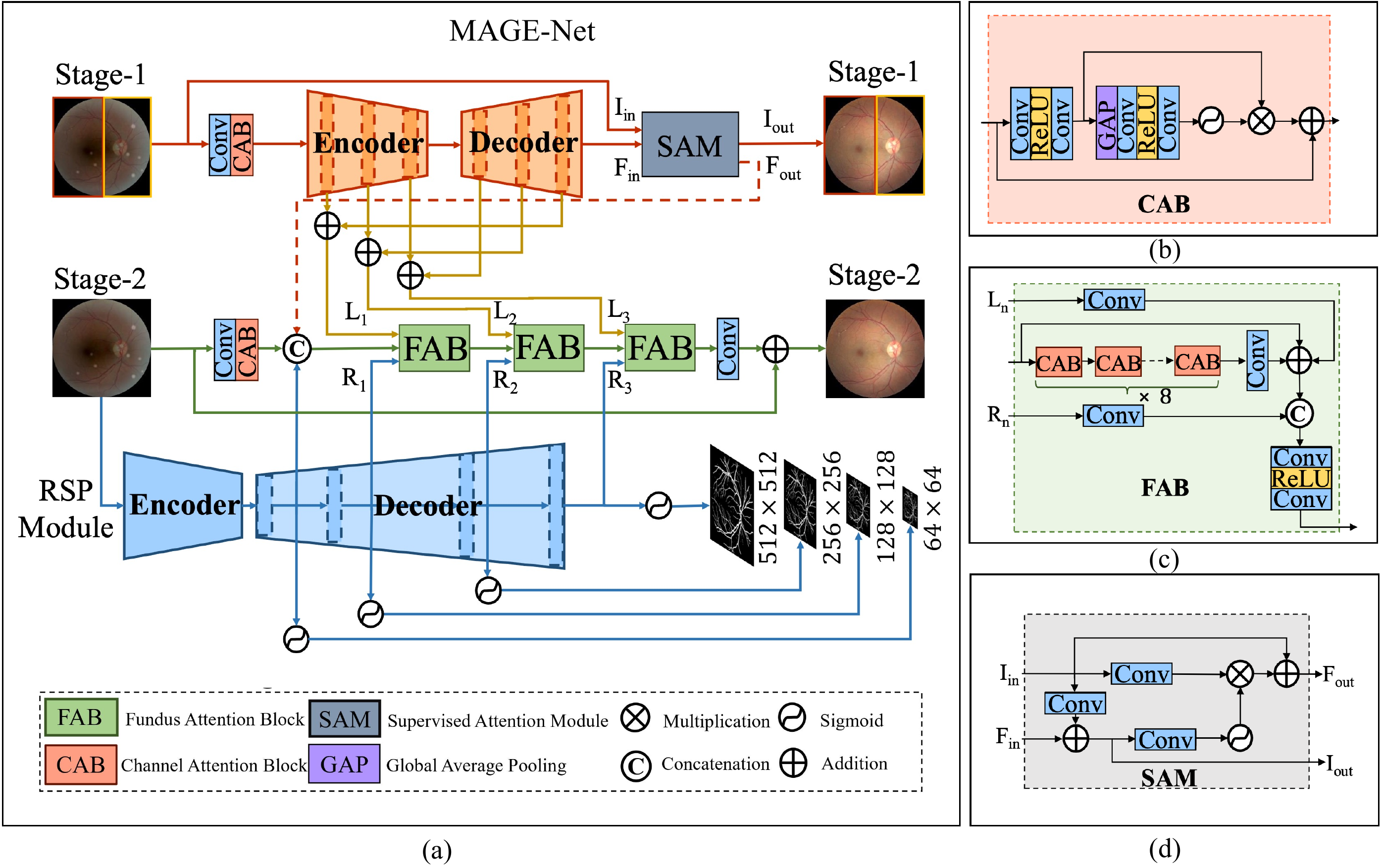}
\end{center}
\caption{\textbf{Illustration of our proposed Multi-Attention
Guided Network (MAGE-Net).} MAGE-Net consists of two parts: multi-stage enhancement module (MSE) and retinal structure preservation module (RSP). For both teacher and student networks, given an input image, we simultaneously feed it into the RSP module and the two stages of the MSE module. The RSP module sends the features maps of important retinal structures to correct stage-2 of the MSE module at each FAB. The stage-1 of the MSE module extracts the contextualized features by a UNet-shaped network. Then, both the encoder-decoder and SAM features are fused into stage-2 from stage-1. (a) Multi-Attention Guided Network (MAGE-Net). (b) Channel Attention Block. (c) Fundus Attention Block. (d) Supervised Attention Module.}\label{fig2}
\end{figure*}

The overview of our proposed teacher-student framework is shown in Fig.~\ref{fig1}. It consists of a student network and a teacher network, both built on our proposed MAGE-Net. In order to integrate synthetic and real images, we also design specific consistency
losses to reduce the domain shift. The detailed structure of MAGE-Net is shown in Fig.~\ref{fig2}a. It is composed of two modules: multi-stage enhancement module (Stage-1 and Stage-2) and retinal structure preservation module (RSP). These two modules are effectively fused through a newly designed fundus attention block (See Fig.~\ref{fig2}c).

\subsection{Multi-Attention Guided Enhancement Network (MAGE-Net)}
\subsubsection{Multi-Stage Enhancement (MSE) Module}
 The MSE module (Fig.~\ref{fig2}a) consists of two stages to restore clean fundus images. Stage-1 employs encoder-decoder structure with large receptive fields to extract the contextualized features in the fundus images. However, the downsampling operations in Stage-1 loss spatial details and thus yield over-smoothed results. Therefore, Stage-2 is proposed for three reasons: preserving local image details by operating on the original image resolution, maintaining the anatomical structure by adding the feature from the RSP module, and fusing features through the fundus attention block effectively. Given a low-quality input image, we feed it into each enhancement stage. To enrich the features from Stage-2, we also adopt a multi-patch hierarchy strategy on the input image from Stage-1. We split each input image into two non-overlapping subimages with $50\%$ of the original resolution. Then, we feed them into Stage-2.

In Stage-1, we first utilize a convolution layer and a channel attention block (CAB)~\cite{zhang2018image} (Fig.~\ref{fig2}b) to extract the features from the input. Specifically, the CAB generates different attention maps for each channel-wise feature, making the network focus on more informative features. Then a UNet-shaped~\cite{ronneberger2015u} architecture is adopted as our sub-network to restore low-quality fundus images. Each encoder layer employs CABs to extract high-level semantic features, and each decoder layer uses a bilinear upsampling operation followed by a convolution layer instead of using transposed convolution due to the checkerboard artifacts introduced by transposed convolution~\cite{odena2016deconvolution}. Both the encoder and decoder features are resized and fused with the intermediate feature maps of Stage-2. Moreover, the output of the decoder is sent through the supervised attention (SAM) module (Fig.~\ref{fig2}d)~\cite{zamir2021multi} to provide the attention maps to Stage-2 with the aid of supervision information from the ground-truth high-quality fundus images.

In Stage-2, we employ a residual network. The features of the original image is firstly extracted by a convolution layer and a CAB like in Stage-1, and then sent to a series of our newly proposed blocks: fundus attention block (FAB) (Fig.~\ref{fig2}c). To preserve local image details, the FAB keeps the feature maps at the original image resolution and does not employ any downsampling operation. In each FAB, we first extract high-resolution features by using several CABs. Then, we add the resized contextual features from Stage-1 with those from Stage-2 to refine the feature maps of whole images. To boost the performance of the downstream clinical analysis tasks, the feature maps from the RSP module are resized and concatenated with the feature maps from Stage-2. Finally, the fused feature maps are passed through a learnable non-linear transformation filter to maintain stable and efficient model performance. We sequentially employ three FABs for feature extraction and fusion to generate the residual and then add it with the low-quality input image to produce the final enhancement output.
%Each FAB combines 8 CABs, as shown in Fig.~\ref{fig1} (a). We employ three FABs to preserve the image details and fuse the features from each scale layer of the RSP module decoder and encoder-decoder of stage-1.

\subsubsection{Retinal Structure Preservation (RSP) Module}
Nature image enhancement methods focus on producing visually satisfying results for humans, without necessarily preserving valuable clinical information in the reconstructed images. However, as important clinical diagnosis evidence, the enhanced fundus images should preserve the retinal structures without incorrectly synthesizing the content. Therefore, we propose the RSP module to further guide image enhancement in the MSE module. As illustrated in Fig.~\ref{fig2}a, the RSP module is based on the pre-trained AG-Net~\cite{zhang2019attention}, which is a UNet-shaped segmentation network with its layers respectively supervised by multi-scale segmentation masks. The main retinal structures are encoded as an attention feature map by the decoder at each scale layer. These multi-scale feature maps from different decoder layers are then fused with the MSE module after being scaled up to the original image size. %To fuse the feature maps into stage-2, first, the resized feature maps are concatenated to the features of stage-2. Second, the feature maps are passed via a learnable non-linear transformation filter, shown in fig.~\ref{fig1}(a).%
In this way, the important components of fundus images are injected into the image correction network to preserve the clinically useful contents.

\subsubsection{MAGE-Net Loss}
%We process the high-quality image $\mathbf {I}_{HQ}$ by the degrade model~\cite{shen2020modeling} to get a synthetic low-quality image. Then, we pass the synthesized low-quality image into our MAGE-Net. At a given stage $S$ of the MSE model, we add the enhanced image with the low-quality input image as the final output $\mathbf {I}_{S}$. At a given layer $V$ of the decoder in the RSP model, We obtain the segmentation output image $\mathbf {I}_{V}$ from the RSP Module. The supervised loss $\mathcal{L}^s$ is defined as the sum of enhancement losses $\mathcal{L}_e$ (the sum of Charbonnier loss~\cite{charbonnier1994two} and the Edge loss~\cite{zamir2021multi}) and segmentation losses $\mathcal{L}_{2}$ loss:
Our MAGE-Net is supervised by an enhancement loss and a segmentation loss. Given a labeled synthetic low-quality image $\mathbf{I}_l$, denoting its enhancement output at the $s$-th stage as $\mathbf{I}_e^s$ and the ground-truth high-quality image as $\mathbf{I}_{h}$, the enhancement loss at the $s$-th stage is the sum of the Charbonnier loss~\cite{charbonnier1994two}:
\begin{equation}
  \mathcal{L}_{char}^s({\mathbf I}_l)=\sqrt{\Vert{\mathbf {I}_{e}^s-\mathbf {I}_{h}}\Vert^{2}+\varepsilon^{2}}
\end{equation}
and the Edge loss~\cite{zamir2021multi}:
\begin{equation}
    \mathcal{L}_{edge}^s({\mathbf I}_l)=\sqrt{\Vert\mathbf{\Delta}(\mathbf {I}_{e}^s) - \mathbf{\Delta}(\mathbf {I}_{h})\Vert^{2}+\varepsilon^{2}},
\end{equation}
where $\varepsilon$ is set as 0.001 in both loss functions, and $\mathbf{\Delta}(\cdot)$ is the gradient function. 
Denoting the segmentation result of $\mathbf{I}_l$ at the $v$-th scale in the RSP module as $\mathbf {I}_{seg}^v$ and the ground-truth mask as $\mathbf{G}_{seg}^v$, the segmentation loss at the $v$-th scale is calculated as:
\begin{equation}
\mathcal{L}_{seg}^v(\mathbf{I}_l)=\Vert\mathbf{I}_{seg}^v-\mathbf{G}_{seg}^v\Vert_{2}
\end{equation}
for each of the four scales in the RSP module. The overall supervised loss of our MAGE-Net is provided below, where the trade-off coefficient $\lambda$ is set as 0.5:
\begin{equation}
\mathcal{L}_{mage}(\mathbf{I}_l)=\sum_{s=1}^{2}(\mathcal{L}^{s}_{char}(\mathbf{I}_l) +  \mathcal{L}^{s}_{edge}(\mathbf{I}_l))+ \lambda \sum_{v=1}^{4} \mathcal{L}^{v}_{seg}(\mathbf{I}_l),
\end{equation}
where $s$ denotes the stage index of MSE module, and $v$ denotes the scale index of the RSP module. 
\iffalse
The enhancement loss $\mathcal{L}_e$ is the sum of Charbonnier loss~\cite{charbonnier1994two} and the Edge loss~\cite{zamir2021multi}:
\begin{equation}
\mathcal{L}_{e}=\mathcal{L}_{char}(\mathbf {I}_{S},\mathbf {I}_{HQ}) +  \mathcal{L}_{edge}(\mathbf {I}_{S},\mathbf {I}_{HQ}),
\end{equation}
%and $\mathcal{L}_{char}$ is the Charbonnier loss~\cite{charbonnier1994two}:
%
%\begin{equation}
%\mathcal{L}_{char}(\mathbf {I}_{S},\mathbf {I}_{HQ})=\sqrt{\Vert{\mathbf %{I}_{S}-\mathbf {I}_{HQ}}\Vert^{2}+\varepsilon^{2}},
%\end{equation}
%
%$\varepsilon$ is equal to 0.001 in this loss function.
%Moreover, the $\mathcal{L}_{edge}$ is the edge loss~\cite{zamir2021multi}:
%\begin{equation}
%\mathcal{L}_{edge}(\mathbf {I}_{S},\mathbf %{I}_{HQ})=\sqrt{\Vert\mathbf{\Delta}(\mathbf {I}_{S}) - %\mathbf{\Delta}(\mathbf {I}_{HQ})\Vert^{2}+\varepsilon^{2}},
%\end{equation}
%Where $\mathbf{\Delta}$ represents the Laplacian operator.
The segmentation losses $\mathcal{L}_{str}$ is L2 loss: 
\begin{equation}
\mathcal{L}_{str}(\mathbf{I}_{V},\mathbf {G}_{V})=\Vert\mathbf{I}_{V}-\mathbf{G}_{V}\Vert_{L_2}
\end{equation}
\fi

\subsection{Transferable MAGE-Net with Multiple Consistency Losses}
Fig.~\ref{fig1} depicts the overall architecture of our proposed Transferable MAGE-Net (T-MAGE-Net) with multiple consistency losses, which corrects the low-quality fundus images for clinical observation and leverages real images for fundus images enhancement. We use the MAGE-Net for both the teacher and student networks. During the training process, we feed the synthetic image pairs into the student network and compute the supervised loss of our MAGE-Net. Meanwhile, for each unlabeled real image without the corresponding enhanced ground-truth, we create an auxiliary image from it by adding Gaussian noise, separately feeding them into the student and teacher networks, and enforce the consistent prediction results from the two networks. The consistency loss is computed for the unlabeled real images at each enhancement stage and each scale of the RSP decoder outputs between the teacher and student networks. The whole architecture is then optimized with two consistency regularizations: 1) each stage of the enhancement results consistency to align the clean outputs predictions between teacher and student, 2) each scale of the segmentation consistency for matching and preserving the retinal structure between teacher and student networks. The weights of the teacher model are updated by the exponential moving average weights of the student model~\cite{tarvainen2017mean}, which will not significantly increase burden to our MAGE-Net becauses of shared weights between teacher and student models. The employment of a teacher-student framework does not introduce additional parameters to learn as
the teacher is simply the exponential moving average of the student~\cite{yang2022unbox}.

%During the training process, we feed the labeled synthetic labeled data into the student network and compute the MAGE-Net loss signed as supervised loss. We create an auxiliary image from unlabeled real images in pairs by adding Gaussian noise and separately feeding them into the student and teacher networks. The consistency loss is computed for unlabeled data on each enhancement stage and each scale of RSP decoder outputs between teacher network and student network. The weights of the teacher model are updated by the exponential moving average weights of the student model.

\subsubsection{Consistency Loss}
%For the real low-quality data, we feed it into the student networks to get the enhanced image $\mathbf{S}_{S}$ from the $S$-th stage, and the segmentation result $\mathbf{S}_{V}$ from the $V$-th layer. At the same time, we add Gaussian noise to the real low-quality fundus image. Then, we feed the noised image into the teacher network to obtain enhanced image $\mathbf{T}_{S}$ from the $S$-th stage, and the segmentation result $\mathbf{T}_{V}$ from the $V$-th layer. Second, we enforce the predictions results of both teacher and student networks to be consistent. Our framework proposes a new consistency loss derived from tow tasks: the enhancement result consistency and segmentation result consistency.
%~~~We propose a new consistency method based on our MAGE-Net. We feed real low-quality images added with Gaussian noise into the student network and teacher network, separately. Then, for the teacher and student network, we enforce the enhancement results ($T_e$ and $S_e$) and segmentation results($T_{seg}$ and $S_{seg}$) to be consistent by the sum of  multi-stage enhancement consistency term and the multi-level segmentation consistency term. Therefore, the multiple consistency $\mathcal{L}^c$ is:
Denote an unlabeled real image as $\mathbf{I}_u$ and its augment as $\tilde{\mathbf{I}}_u$, and let $S(\cdot)$ and $T(\cdot)$ represent the embedding functions of the student and the teacher networks, respectively. We enforce the two networks to output consistent enhancement results ($S_e(\mathbf{I}_u)$ and $T_e(\tilde{\mathbf{I}}_u)$) of each stages and segmentation results($S_{seg}(\mathbf{I}_u)$ and $T_{seg}(\tilde{\mathbf{I}}_u)$) of each segmentation scale.  The overall consistency loss $\mathcal{L}_{cons}(\mathbf{I}_u)$ sums over the multi-stage enhancement consistency loss and the multi-level segmentation consistency losses: 
\begin{equation}
\begin{split}
\mathcal{L}_{cons}(\mathbf{I}_u)=\sum_{s=1}^{2}\Vert{S}^{s}_{e}(\mathbf{I}_u) - {T}^{s}_{e}(\tilde{\mathbf{I}}_u)\Vert_{1} \\ 
+ \sum_{v=1}^{4}\Vert{S}^{v}_{seg}(\mathbf{I}_u) - {T}^{v}_{seg}(\tilde{\mathbf{I}}_u)\Vert_{1}.
\end{split}
\end{equation}

where $s$ denotes the stage index of MSE module, and $v$ denotes the scale index of the RSP module. 
%\subsection{Overall Loss of MAGE-Net+Teacher-Student Framework}
\subsubsection{Total Loss}
The total loss of our method is the sum of the supervised loss from our MAGE-Net and the unsupervised multi-stage multi-level consistency loss: 
\begin{equation}
\mathcal{L}=\sum^{M}_{i=1}\mathcal{L}_{mage}(\mathbf{I}_l^i) +\mu \sum^{N}_{j=1}\mathcal{L}_{cons}(\mathbf{I}_u^j),
\end{equation}
where $M$ and $N$ denote the total number of the labeled and unlabeled images, respectively. The weight $\mu$ is computed by a time-dependent Gaussian warming up function~\cite{tarvainen2017mean}. %$\mu(t)=\mu_{max}e^{(-5(1-t/t_{max})^2)}$
%, where $t$ denotes the current training iteration and $t_{max}$ is the maximum training iteration, and we set $t_{max} = 1$. The parameters of the student network are updated to minimize $\mathcal{L}$.
The parameters of teacher network are updated by the exponential moving average (EMA) strategy in each training iteration.

%====================================%
\section{Experiments}
\subsection{Implementation Details}
Our method is implemented by PyTorch, and trained on a single NVIDIA RTX V6000 GPU. The Adam optimizer is adopted. The initial learning rate is $ 2\times {10}^{-5} $, which is decreased to $ 1\times {10}^{-7} $ by the cosine annealing strategy~\cite{loshchilov2016sgdr}. All of the labeled and unlabeled images are re-scaled to the size of $ 512 \times 512 $. The mini-batch size is 24, including 16 labeled synthetic images and 8 unlabeled real images. We use a two-stage training strategy. In order to accelerate our training process, we pretrain the RSP module, and then  train the whole enhancement framework in an end-to-end fashion.

\subsection{Datasets}
Our training set is formed from the EyeQ~\cite{fu2019evaluation} dataset, whose images are captured by various cameras from different hospitals. From this perspective,  our model is generally trained to enhance
images from different centers or equipments~\cite{nan2022data}. The EyeQ~\cite{fu2019evaluation} dataset is a subset of the Kaggle~\cite{kaggle} dataset for fundus image quality assessment, which has 28,792 retinal images with three quality grades (“Good”, “Usable”, and “Reject”). Specifically, we select 10,000 high-quality fundus images (labeled as “Good”) as the clean images and produce segmentation masks by using pre-trained AG-Net~\cite{zhang2019attention}. We randomly choose degradation factors (e.g., light transmission disturbance, image blurring, and retinal artifacts) to synthesize degraded images by using the method~\cite{shen2020modeling}. Moreover, we randomly select 5,000 low-quality images (labeled as “Uable”) as the unlabeled data. In the test stage, to evaluate the quality of image enhancement, we also utilize the degradation model~\cite{shen2020modeling} to randomly generate degraded
images on the DRIVE~\cite{staal2004ridge} test set and REFUGE~\cite{orlando2020refuge} test set. Moreover, For the Subtest-EyeQ dataset, we chose another 500 images which are labeled as “Good” in EyeQ but not present in Subtrain-EyeQ dataset to evaluate the image enhancement quality.

\subsection{Degradation Model Settings}
The degradation method is based on the ophthalmoscope imaging systems, which is also verified by Cofe-Net~\cite{shen2020modeling}. Clinical image
collection in a complex environment using an ophthalmoscope often encounters several types of interference, as introduced in the
optical feed-forward system. Light transmission disturbance is often caused by exposure issues. Due to the interspace between
the eye and camera, stray light may enter into the ophthalmoscope, mix with the lighting source and result in uneven exposure.
This also affects the tuning setting of the programmed exposure, leading to global over-/under-exposure. In addition, image
blurring caused by human factors (such as eyeball movement, fluttering, and defocus) results in low-quality images. Besides,
the capturing of undesired objects (e.g., dust) during imaging is also a crucial factor that reduces image quality and impedes
subsequent diagnosis. Therefore, Cofe-Net~\cite{shen2020modeling} proposes a reformulated representation of the interference that occurs during the
collection of fundus images. The degradation model could be used to not only support current fundus propagation models,
but also synthesize a high-quality pairwise fundus dataset for subsequent research. Cofe-Net ~\cite{shen2020modeling} summarized the interference in
terms of three factors, including light transmission disturbance, image blurring, and retinal artifacts. Thanks to Cofe-Net ~\cite{shen2020modeling},
the degradation method was directly adopted in subsequent works, such as I-SECRET~\cite{cheng2021secret}. We put a high-quality fundus image $\mathbf{x}$ into the degradation model to get the paired degraded image $x^{\prime}$.

\textbf{Light Transmission Disturbance.}: The light transmission disturbance contains two types of degraded factors: global factors and local factors. The global factors include contrast factor, brightness, and saturation, which are caused by unstable stray light, subjective situation, and manual mydriasis. The local factors produce additional non-uniform illumination due to the initiative light leak phenomenon, diverse lens apertures, and embedded optical compensation mechanism. Therefore, light transmission disturbance is simulated by using:
\begin{equation}
\mathbf{x^{\prime}}=clip(\alpha(\mathbf{J}\cdot G_{L}(r_{L},\sigma_{L})+\mathbf{x})+\beta;s),
\end{equation}
where $\alpha$, $\beta$, and $s$ refer to the contrast factor, brightness, and saturation, respectively. To simulate global factors, we randomly set them between $-0.5$ to $0.5$. $G_{L}$ is a Gaussian filter with the radius $r_L$ and the variance $\sigma_L$. For local factors, an illumination bias $\mathbf{J}$ is defined as:
\begin{equation}
\mathbf{J}_{ij} = n_{l}\mid_{(i-a)^{2}+(j-b)^{2}<r^2_L},
\end{equation}
where $c=(a, b)$ is the center with the radius of $r_{L}$. We randomly set $c\in[0.375r_{L}, 0.625r_{L}]$. We define $r_{L}\in[0.75w, w]$; $\sigma_{L}\in[0.66cr_{L}, 0.66(w-c)r_{L}]$ and $r_{L}\in[0.3w, 0.5w]$; $\sigma_{L}\in[0.55r_L, 0.75r_L]$ for light leak phenomenon and uneven exposure problem, respectively, where $w$ denotes the image size.

\textbf{Image Blurring.}: The image blurring is caused by undesired object distance in funduscopy. It is simulated by using:
\begin{equation}
\mathbf{x^{\prime}}=\mathbf{x}\cdot{G_{B}}(r_{B},\sigma_{B})+n,
\end{equation}
where $G_{B}$ is a Gaussian filter with a radius $r_{B}$ and the spatial
constant $\sigma_{B}$, and $n$ denotes the additive random Gaussian
noise. Here we set $\sigma_{B}=0.03w$, and $r_{B}\in[0.01w, 0.015w]$.

\textbf{Retinal Artifact.}: The retinal artifacts are caused by dust and grains attaching on
the lens of the imaging plane. It is simulated by using:
\begin{equation}
\mathbf{x^{\prime}}=\mathbf{x}+\sum_{k}^{K}{G_{R}}(r_{k}/4,\sigma_{k})\cdot \mathbf{o_{k}},
\end{equation}
where $K$ is the undesired object number. To simulate the interference in real clinical scenarios, we randomly increase the number of undesired objects from 10 to 30. For each undesired object $k$, $r_{k}$ and $\sigma_{k}$ are defined as the radius and the variance of a Gaussian filter $G_{R}$ . We randomly set the radius $r_{k}\in[0.025w, 0.05w]$, the variance $\sigma_{k}=5+0.8r_{k}$, and the illumination bias $\mathbf{o_{k}}=1-e^{-(0.5+0.04r_{k})\times(0.012r_{k})}$ for each object $k$.

%----------------------%
%
\begin{table}[!t]
\centering
\footnotesize
\caption{Image enhancement quality comparison of different methods on DRIVE~\cite{staal2004ridge}, REFUGE~\cite{orlando2020refuge}, and Subtest-EyeQ}.\label{tab1}
\begin{tabular}{|l|cc|cc|cc|}
\hline
 &\multicolumn{2}{c|}{DRIVE~\cite{staal2004ridge}} & \multicolumn{2}{c|}{REFUGE~\cite{orlando2020refuge}}&\multicolumn{2}{c|}{Subtest-EyeQ}\\
\textbf{Methods}&\textbf{PSNR}&\textbf{SSIM}&\textbf{PSNR}&\textbf{SSIM}&\textbf{PSNR}&\textbf{SSIM}\\
\hline
Low-quality &16.28 &0.804 &16.79 &0.823&18.49&0.811\\
Setiawan~\cite{setiawan2013color}&16.68 &0.680 &16.65 &0.668&18.59&0.770\\
DCP~\cite{he2010single} &17.55 &0.792 &14.66 &0.702&17.98&0.786\\
Cofe-Net~\cite{shen2020modeling}& 20.31 & 0.881 &24.45 &0.897 &21.88&0.880\\
StillGAN~\cite{ma2021structure}& 22.48 & 0.890 &22.90 &0.871&22.77&0.850\\
I-SECRET~\cite{cheng2021secret}&24.09  & 0.906 &25.04&0.914&23.53&0.889\\
\textbf{T-MAGE-Net} &\bf24.72& \bf0.928 &\bf 25.66&\bf 0.929&\bf{24.14}&\bf{0.896}\\
\hline
\end{tabular}
\end{table}

\begin{table}[!t]
\centering
\caption{Comparison of different methods in terms of the number of parameters.}\label{tab7}
\begin{tabular}{|l|c|}
\hline
\textbf{Methods}&\textbf{Number of Parameters (M)}\\
\hline

Cofe-Net~\cite{shen2020modeling}& 41.218\\
StillGAN~\cite{ma2021structure}& 78.644\\
I-SECRET~\cite{cheng2021secret}& 11.756\\
T-MAGE-Net(ours) & 26.379\\
\hline
\end{tabular}
\end{table}

\begin{table}[!t]
\centering
\caption{Vessel segmentation performance comparison of different methods on DRIVE~\cite{staal2004ridge} dataset.}\label{tab2}
\begin{tabular}{|l|ccc|}
\hline
\textbf{Methods}&\textbf{AUC}&\textbf{Acc}&\textbf{IoU}\\
\hline
Low-quality &0.781 &0.938 &0.479\\
high-quality &0.953&0.982&0.830\\
Setiawan~\cite{setiawan2013color}&0.809 &0.938 &0.504\\
DCP~\cite{he2010single}&0.813 &0.948 &0.547\\
Cofe-Net~\cite{shen2020modeling}& 0.875 & 0.961 & 0.654\\
StillGAN~\cite{ma2021structure}& 0.867 & 0.959 & 0.639\\
I-SECRET~\cite{cheng2021secret}& 0.877 & 0.963 &  0.662\\
MAGE-Net(ours)&0.910&0.970&0.726\\
\textbf{T-MAGE-Net(ours)} &\bf 0.932 & \bf 0.974 & \bf 0.764 \\
\hline
\end{tabular}
\end{table}

\section{Evaluation}
In this section, we evaluate the enhancement performance of different methods in terms of the image quality and three downstream tasks including vessel segmentation task, optic disc/cup detection task and real clinical image analysis task.

\subsection{Image Quality Enhancement}
For quantitative evaluation, we use both PSNR and SSIM as the evaluation metrics in Table~\ref{tab1}. Our method is the best performer in terms of both PSNR and SSIM~\cite{wang2004image}. Specifically, among all comparing methods, the baseline method proposed by Setiawan et al.~\cite{setiawan2013color} and the DCP~\cite{he2010single} method correct each fundus image based on the global image statistics and functions, so their results contain undesired distortion; The relatively poor work results from both methods Setiawan et al.~\cite{setiawan2013color} and DCP~\cite{he2010single} are expected since they are non-deep learning methods. The method from Setiawan et al.~\cite{setiawan2013color} exploits the image contrast normalization and contrast limited adaptive histogram equalization (CLAHE) techniques to restore the color of retinal images. Instead of simply considering the color and texture information like Setiawan et al.~\cite{setiawan2013color}, the DCP~\cite{he2010single} method decomposes the reflection and illumination, which achieves image enhancement and correction by estimating the solution through an alternative minimization scheme. While these algorithms based on the bottom-up frameworks are effective, their optimal solutions rely heavily on global image statistics and mapping functions, namely, these methods ignore discriminative features, which may introduce undesired artifacts and distortion. StillGAN~\cite{ma2021structure}, a CycleGAN-like method, is trained with unpaired supervision that could not well preserve the local structure details; Cofe-Net~\cite{shen2020modeling} model trained on synthetic image pairs ignores the gap between the synthetic and the real low-quality images for the practical diagnosis, so it is not well generalized to the authentic clinical fundus images.  Although utilizing unlabelled real fundus images, I-SECRET~\cite{cheng2021secret} method still loses to ours as it only uses a single CNN-based network to enhance the fundus images without well preserving the retinal and lesion details. Moreover, by cross-referencing Table~\ref{tab5}, it is found that our method still wins all the baseline methods even without using the RSP module (i.e., without the additional guidance from the extra segmentation masks), since our method without RSP (i.e., “Ours w/o RSP”) achieves PSNR/SSIM of 24.47/0.9273, better than the best baseline I-SECRET (24.09/0.906) shown in Table~\ref{tab1}. This may suggest that our mean teacher based domain adaptation more effectively utilizes the unlabelled real low-quality images, compared with the contrastive learning employed by I-SECRET. %applies the adversarial loss to achieve more photo-realistic results from real high-quality images.However, due to the unbalance scenario between the high-quality health and illness images, the I-SECRET~\cite{cheng2021secret} method cannot well preserve the lesion area preservation.
%Therefore, our method is more suitable for the practical clinical application.
%Our method consistently achieves better results than other competitors. 

\begin{figure*}[!t]
\includegraphics[width=\textwidth]{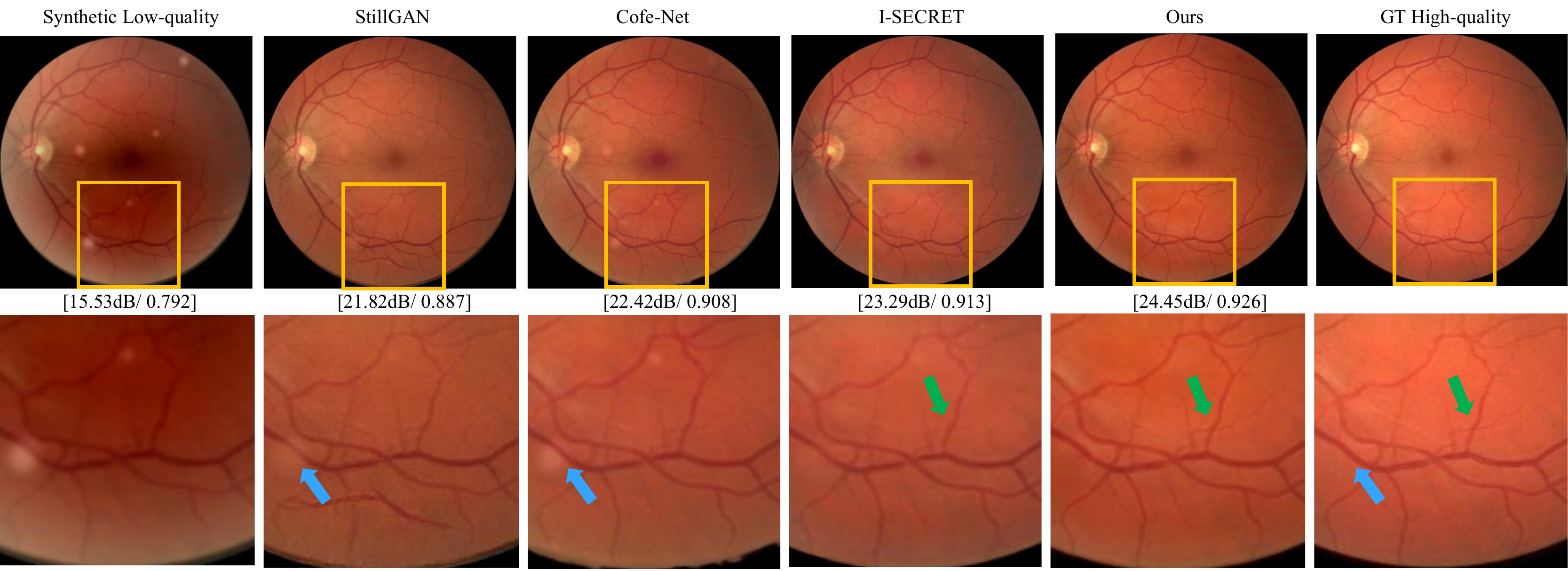}
\caption{\textbf{Visual comparison of enhancement results (Columns 2-5) on a synthetic %(the top row) and the real (the bottom row) 
low-quality fundus image (Column 1)}. The ground-truth (GT) high-quality image is given in Column 6. The symbol [. / .] denotes [PSNR / SSIM] scores. The bottom row contains the zoom-in views of the images in the top row. Arrows point to the visual differences for attention. 
%For the StillGAN~\cite{ma2021structure}, it produces fake vessels compared with the high-quality images. For the StillGAN~\cite{ma2021structure} and the Cofe-Net~\cite{cheng2021secret}, the undesired light spot presented in the low-quality synthetic image still remain in the enhanced images. For  I-SECRET~\cite{cheng2021secret}, the tiny vessel structures are blurred compared with our results. 
%%The high-quality image, synthetic low-qualit image, and enhancement results (the top row). Real low-qualit image, and enhancement results (the bottom row).
} \label{fig3} 
\end{figure*}

\begin{figure}[!t]
\centerline{\includegraphics[width=\columnwidth]{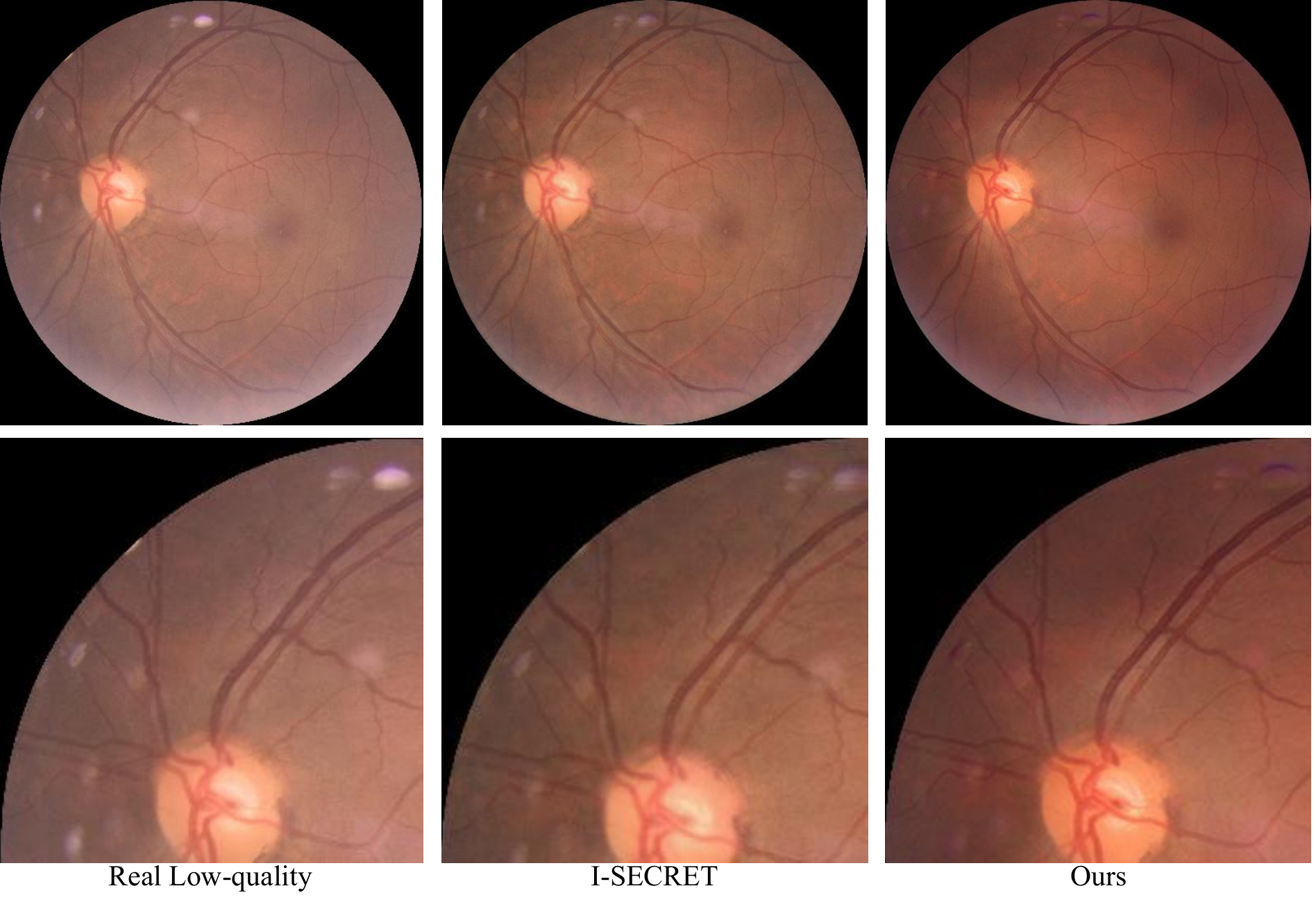}}
\caption{Visual comparison between I-SECRET~\cite{cheng2021secret} and our method by enhancement from  a real low-quality fundus image. The top row shows the real low-quality fundus image and the enhancement results. The bottom row shows the zoom-in views.}
\label{fig0}
\end{figure}

\begin{figure*}[!t]
\begin{center}
\includegraphics[width=0.9\textwidth]{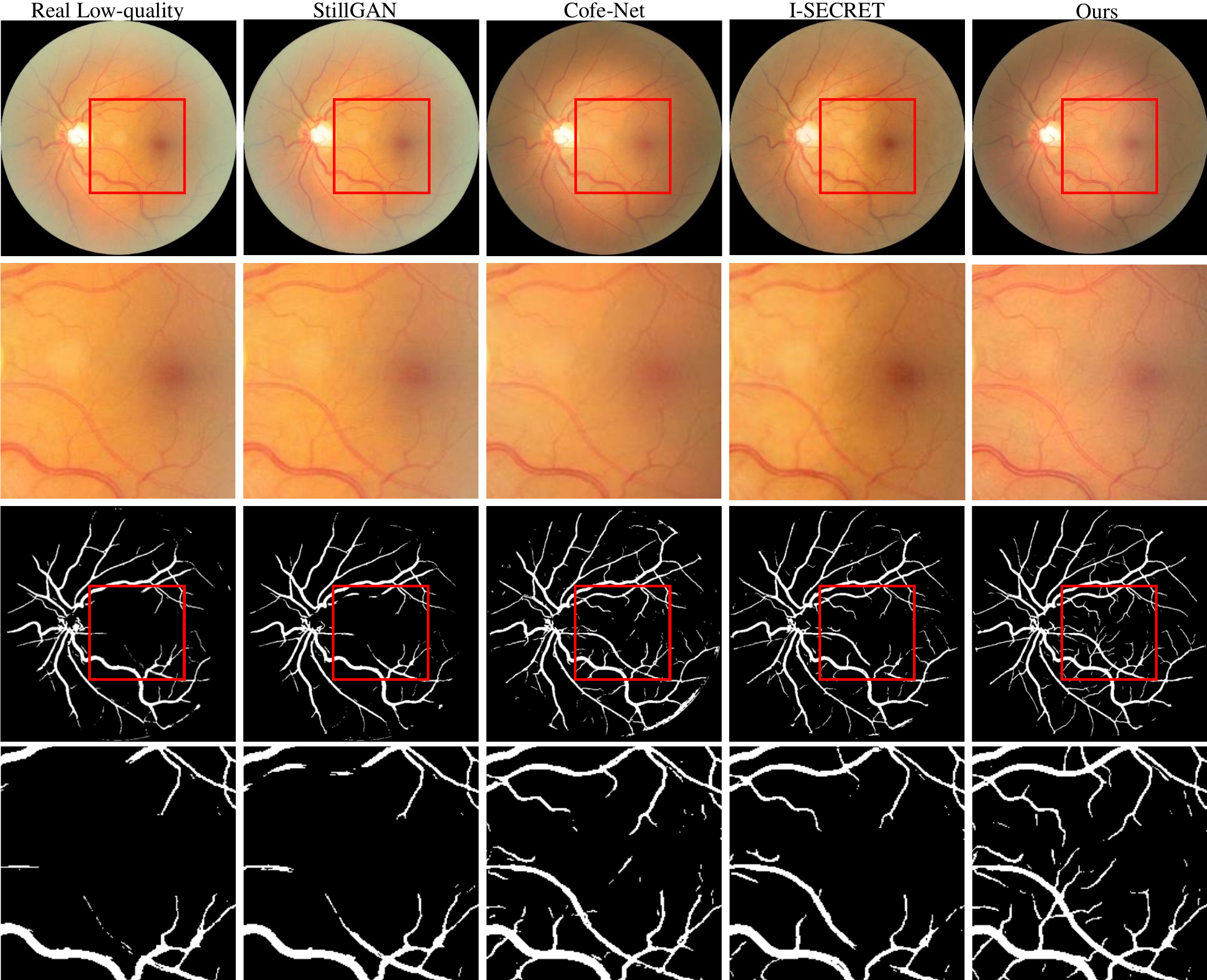}
\end{center}
\caption{\textbf{Visual comparison of enhancement results on real low-quality fundus images.} From top to bottom the images are the enhanced images and their zoom-in, and the vessel segmentation results and their zoom-in. The real low-quality images are given in the 1st column. There are no ground-truth high-quality images.}
\label{fig4}
\end{figure*}

In addition, visual comparisons of different enhancement results are given in Fig.~\ref{fig3} for synthetic low-quality images and in Fig.~\ref{fig0} and Fig.~\ref{fig4} for real low-quality images without ground-truth. From Fig.~\ref{fig3}, we can see that neither StillGAN nor Cofe-Net could eliminate the undesired light spot (indicated by the blue arrows) presented in the synthetic low-quality image. This problem is alleviated in the images enhanced by I-SECRET, as it considers both labeled and unlabeled retina images to learn robust features. However, compared with our method, I-SECRET generates blurred vessel boundaries (indicated by the green arrows), confirming the importance of introducing RSP module. Our advantage over I-SECRET could also be observed by the enhancement result from a real low-quality image in Fig.~\ref{fig0}. Compared with I-SECRET, our method produces a sharper image with much less undesired light spots. Visual comparison with more methods on real low-quality images is given in Fig.~\ref{fig4}. Since there is no ground-truth for the enhancement, we additionally show the vessel segmentation from the enhanced images. As shown, our method could recover finer vessels than other methods.
We compared the number of training parameters between the proposed T-MAGE-Net and other baseline approaches, as shown in Table~\ref{tab7}. It could be found that the numbers of parameters of Cofe-Net~\cite{shen2020modeling} and StillGAN~\cite{ma2021structure} are much more than that of our method. Ours is just a bit more than that of I-SECRET~\cite{cheng2021secret}, but ours achieves better performance.

\subsection{Vessel Segmentation}
The enhancement quality is also validated through the downstream task of vessel segmentation. We evaluate different methods on the degraded DRIVE~\cite{staal2004ridge} test set. Quantitative results are shown in Table~\ref{tab1}. Our method outperforms the competing methods in terms of AUC, Accuracy (Acc.), and IoU. We use CE-Net\cite{gu2019net} trained on DRIVE training set as the segmentation method, which achieves AUC/Acc/IoU of 0.953/0.982/0.830 on high-quality DRIVE~\cite{staal2004ridge} test set. The vessel segmentation results of real low-quality fundus images from different deep learning methods are shown in Fig.~\ref{fig4}. Obviously, the vessel structure is better preserved by using our enhancement method. In contrast, the baseline method by Setiawan et al.~\cite{setiawan2013color} and DCP~\cite{he2010single} produce unsatisfactory results, because their solutions highly rely on global image contrast and illumination. They cannot correct local light spots, holes, and halos, which influences the vessel observation. Other baseline methods, like StillGAN~\cite{ma2021structure} and I-SECRET~\cite{cheng2021secret}, use different supervised losses to preserve the structure information of the whole images, but the useful retinal structures for clinic diagnosis are not emphasized during the process, so that the vessel details are missed in Fig.~\ref{fig4}. Differently, Cofe-Net~\cite{shen2020modeling} designed a retinal structure activation module to emphasize the anatomical retinal structures. Comparing with it, our RSP module together with a multi-resolution method can provide more effective and robust structural features. Moreover, using the restored images from our MAGE-Net without RSP for vessel segmentation, we achieve the results of AUC/Acc/IoU as 0.910/0.970/0.726, better than I-SECRET's 0.877/0.963/0.662 as reported in Table~\ref{tab2}. The results indicate that our improvements come from both the RSP module to preserve retinal structure and the MSE module to achieve clear images. Therefore, our enhanced images with better quality also lead to the smallest error of vessel segmentation over those produced by other baseline methods. 

\subsection{Optic Disc/Cup Detection}
Optic disc and cup detection is important for diagnosing glaucoma. For the downstream task of optic disc/cup detection, we evaluate different methods on the degraded REFUGE~\cite{orlando2020refuge} test set. We report the Dice and mIoU in Table~\ref{tab3}. Specifically, we use Pra-Net~\cite{fan2020pranet} trained on the REFUGE training set(obtains mIoU/Dice of 0.789/0.882) for the detection on the enhanced images. The results consistently show that the enhanced images by our method could benefit the optic disc and cup detection for clinical observation. By performing paired t-tests based on the optic disc/cup detection results from our method and the best baseline (I-SECRET~\cite{cheng2021secret}), we achieve the p-values of 0.00157 (mIoU) and 0.00063 (Dice), indicating our improvements are statistically significant, as both p-values are lower than 0.05. 
%The results indicate again that our method achieves clearer images than the other methods. 

\begin{table}[!t]
\centering
\caption{Optic disc/cup segmentation performance comparison of different methods on REFUGE~\cite{orlando2020refuge} dataset.}\label{tab3}
\begin{tabular}{|l|cc|}
\hline
\textbf{Methods}&\textbf{mIoU}&\textbf{Dice}\\
\hline
Low-quality &0.709 &0.823\\
high-quality &0.789 &0.882\\
Setiawan~\cite{setiawan2013color} &0.727 &0.841\\
DCP~\cite{he2010single} &0.720 &0.831\\
Cofe-Net~\cite{shen2020modeling} & 0.758 &0.858\\
StillGAN~\cite{ma2021structure} &0.725 &0.834\\
I-SECRET~\cite{cheng2021secret} &0.750 &0.852\\
\textbf{T-MAGE-Net(ours)} & \bf 0.762 &\bf 0.862\\
\hline
\end{tabular}
\end{table}

\begin{table*}[!t]
\centering
\caption{The ocular disease recognition results of each disease on the ODIR~\cite{ODIR-5K} dataset.}\label{tab4}
\resizebox{\textwidth}{!}{
\begin{tabular}{|l|ccc|ccc|ccc|ccc|ccc|ccc|ccc|ccc|ccc|ccc|}
\hline
 &\multicolumn{3}{c|}{normal} & \multicolumn{3}{c|}{diabetes} & \multicolumn{3}{c|}{glaucoma} & \multicolumn{3}{c|}{cataract}& \multicolumn{3}{c|}{AMD} & \multicolumn{3}{c|}{hypertension}
& \multicolumn{3}{c|}{myopia} & \multicolumn{3}{c|}{other diseases} & \multicolumn{3}{c|}{average}\\
\textbf{Methods}& {Kappa}& {ACC}& {AUC}
& {Kappa}&  {ACC}& {AUC}& {Kappa}& {ACC}& {AUC}& {Kappa}&  {ACC}& {AUC} & {Kappa}&  \textbf{ACC}& {AUC}& {Kappa}& {ACC}& {AUC}& {Kappa}&  {ACC}& {AUC}& {Kappa}& {ACC}& {AUC}& {Kappa}&  {ACC}& {AUC}\\
\hline
Real fundus image 
&0.2989&0.7875&0.7925
&0.5100&0.8974&0.7926
&0.4986&0.8875&0.8901
&0.8107&0.9625&0.9776
&0.7151&0.9425&0.9596
&0.2142&0.8898&0.7895
&0.4049&0.9100&0.9117
&0.1694&0.7825&0.6595
&0.4438&0.8825&0.8358
\\
Setiawan et al. [5] 
&0.1501&0.7835&0.7835
&0.2258&0.8800&0.7177
&0.5862&\textbf{0.9249}&0.8761
&0.5985&0.9325&0.9013
&0.3181&0.7000&0.9199
&0.0049&0.8725&0.5980
&0.3043&0.9000&\textbf{0.9716}
&0.0040&\textbf{0.8475}&0.4363
&0.2887 &0.8587& 0.7541\\
DCP [40] 
&0.2899&0.7475&0.8100
&0.3671&0.8675&0.7722
&0.6026&0.9247&0.8975
&0.7350&0.9500&0.9427
&0.6270&0.9275&0.9259
&0.3225&\textbf{0.8950}&0.7877
&0.7615&\textbf{0.9550}&0.9437
&0.1572&0.8425&0.6562
&0.4662&0.8887&0.8488\\
Cofe-Net [7] 
&0.2495&0.7275&0.7868
&0.3766&0.8800&0.7771
&0.5405&0.9150&0.8780
&0.7894&0.9575&0.9676
&0.6832&0.9325&0.9590
&0.2781&0.8800&0.8076
&0.4049&0.9100&0.8626
&\textbf{0.2445}&\textbf{0.8475}&0.6348
&0.4328&0.8812&0.8336
\\
StillGAN [11] 
&0.2500&0.7840&0.7515
&0.3591&0.8550&0.7932
&0.5580&0.9025&0.8870
&0.6794&0.9425&0.9817
&\textbf{0.7977}&\textbf{0.9550}&0.9628
&\textbf{0.3533}&0.8948&\textbf{0.8250}
&0.4741&0.9175&0.9514
&0.2363&0.7900&\textbf{0.6945}
&0.4500&0.8828&0.8528
\\
I-SECRET [21] 
&0.2527&0.7875&0.7977
&0.4110&0.8800&0.8074
&0.5934&0.9225&0.8910
&0.7961&0.9600&0.9747
&0.7346&0.9350&0.9713
&0.3529&0.8900&0.8197
&0.5169&0.9220&0.8963
&0.1566&0.8250&0.6658
&0.4707& 0.8903& 0.8523\\
\hline
MAGE-Net(ours) 
&0.2835&0.7600&0.7900
&0.4388&0.8825&\textbf{0.8090}
&\textbf{0.6160}&0.9225&0.8935
&\textbf{0.8527}&\textbf{0.9700}&\textbf{0.9840}
&0.7942&0.9547&\textbf{0.9790}
&0.2440&0.8800&0.8125
&0.4741&0.9175&0.9266
&0.1656&0.8300&0.6672
&0.4772& 0.8897 &0.8526
\\
\textbf{T-MAGE-Net(ours)} 
&\textbf{0.3438}&\textbf{0.7900}&\textbf{0.8107}
&\textbf{0.5147}&\textbf{0.9075}&0.8076
&0.5660&0.9075&\textbf{0.8989}
&0.8390&0.9675&0.9766
&0.7537&0.9475&0.9692
&0.2203&0.8850&0.7758
&\textbf{0.5175}&0.9225&0.9121
&0.2267&0.7975&0.6727
&\textbf{0.4879}&\textbf{0.8906}&\textbf{0.8545}\\
\hline
\end{tabular}}
\end{table*}

\subsection{Real Clinical Image Analysis}
The enhancement model is expected to provide clean images with lesion preservation to assist diagnosis. This is validated by using the ODIR-5K dataset~\cite{ODIR-5K} collected from different hospitals and medical centers with different image qualities, which contains eight different labels including ``normal", ``diabetes", ``glaucoma", ``cataract", ``age-related macular degeneration (AMD)", ``hypertension", ``myopia", and ``other diseases". We adopt the Jordi et al.~\cite{jordi2019ocular} model 
% (trained an inception ResNetV2 model on 6600 images and tested on 400 images in the ODIR-5K dataset) 
to classify ocular diseases. The results are shown in the Table~\ref{tab4}. Our method boosts the disease recognition performance evidently. Due to the gap between the synthetic and the real low-quality images, Cofe-Net~\cite{shen2020modeling} increases the risk of changing lesion areas, such as color distortion in Fig.~\ref{fig3}. The enhanced images by I-SECRET~\cite{cheng2021secret} and StillGAN~\cite{ma2021structure} are over-smoothed and fail to restore the retinal structure details that are important diagnose clue for eye diseases such as age-related macular degeneration, hypertension, and myopia~\cite{soomro2019deep}. In contrast, we preserve the retina structure by our RSP module, and enhance the feature learning of our MAGE-Net based on both synthetic low-quality images (via the supervised enhancement process from the student model) and real low-quality images (via the unsupervised consistent enhancement process from both teacher and student models). In this way, the learned features can cater for both domains, which helps reduce domain gap to some extent. Without the teacher student framework, the results of Kappa/ACC/AUC become 0.4772/0.8897/0.8526, worse than 0.4879/0.8906/0.8545 achieved by our complete model under the teacher student framework, as shown in Table~\ref{tab4}. 

%Therefore, we design the RSP module to preserve lesion areas, and effectively integrate real fundus images by using teacher-student model to boost disease recognition accuracy.
%\begin{table}[!t]
%\centering
%\caption{The ocular disease recognition results on the ODIR dataset~\cite{ODIR-5K}.}\label{tab4}
%\begin{tabular}{|l|ccc|}
%\hline 
%\textbf{Methods}& \textbf{Kappa}&  \textbf{F-1}& \textbf{AUC}\\
%\hline
%Real fundus image & 0.4693 & 0.8906 & 0.8382\\
%Setiawan et al.~\cite{setiawan2013color} & 0.2829 & 0.8563 & 0.7606\\
%DCP~\cite{he2010single} & 0.4704 & 0.8893 &  0.8485\\
%Cofe-Net~\cite{shen2020modeling} & 0.4582 & 0.8872 & 0.8357\\
%StillGAN~\cite{ma2021structure} & 0.4695 & 0.8878 & 0.8598\\
%I-SECRET~\cite{cheng2021secret} & 0.4778 & 0.8903 & 0.8522\\
%MAGE-Net(ours) &0.4786&0.8880&0.8597\\
%\textbf{T-MAGE-Net(ours)} & \textbf{0.4893} & \textbf{0.8934} & \textbf{\bf 0.8630}\\
%\hline
%\end{tabular}
%\end{table}

\subsection{Ablation Study}
To investigate the contribution of each component in our method, the ablation study is reported in the Table~\ref{tab5}. Initially, stage-1 is a UNet-shaped network to correct fundus images, so some fine details are lost due to the sequential downsampling. To solve this problem, stage-2 is introduced to maintain important information. In order to emphasize retinal structures, we utilize the RSP module for clinical purposes. Our MAGE-Net combines the two stages and the RSP module. We observe that either the multi-stage strategy or the RSP module improves the PSNR and SSIM results. The employment of the  teacher-student framework reduces the domain shift between the authentic image pairs and the real clinic images, which further improves both PSNR and SSIM results.

\begin{table}[!t]
\centering
\caption{ The ablation study results on the DRIVE dataset~\cite{staal2004ridge}, where``S1" denotes the Stage-1, ``S2" denotes the Stage-2, ``TS" denotes the Teacher-student framework, ``Le" denotes the supervised enhancement loss, ``Ls" denotes the supervised segmentation loss, ``Lce" denotes the consistent enhancement loss, and ``Lcs" denotes the consistent segmentation loss.}\label{tab5}
\begin{tabular}{|c|cccc|cc|}
\hline
Loss Combination&S1&S2&RSP&TS&\textbf{PSNR}&\textbf{SSIM} \\ 
\hline
Le&\checkmark& & & &22.93 &0.9109\\
Le&\checkmark&\checkmark & & &23.35 &0.9168\\
Le+Ls&\checkmark&\checkmark&\checkmark& &23.72 &0.9228 \\
Le+Lce&\checkmark&\checkmark& &\checkmark &24.47 &0.9273 \\
Le+Ls+Lce+Lcs(w/o CAB)&\checkmark&\checkmark&\checkmark&\checkmark&23.33&0.9212\\
Le+Ls+Lce+Lcs&\checkmark&\checkmark&\checkmark&\checkmark&\bf24.72&\bf0.9281\\
\hline
\end{tabular} 
\end{table}

\subsection{Limitation}
\begin{figure}[!t]
\centerline{\includegraphics[width=\columnwidth]{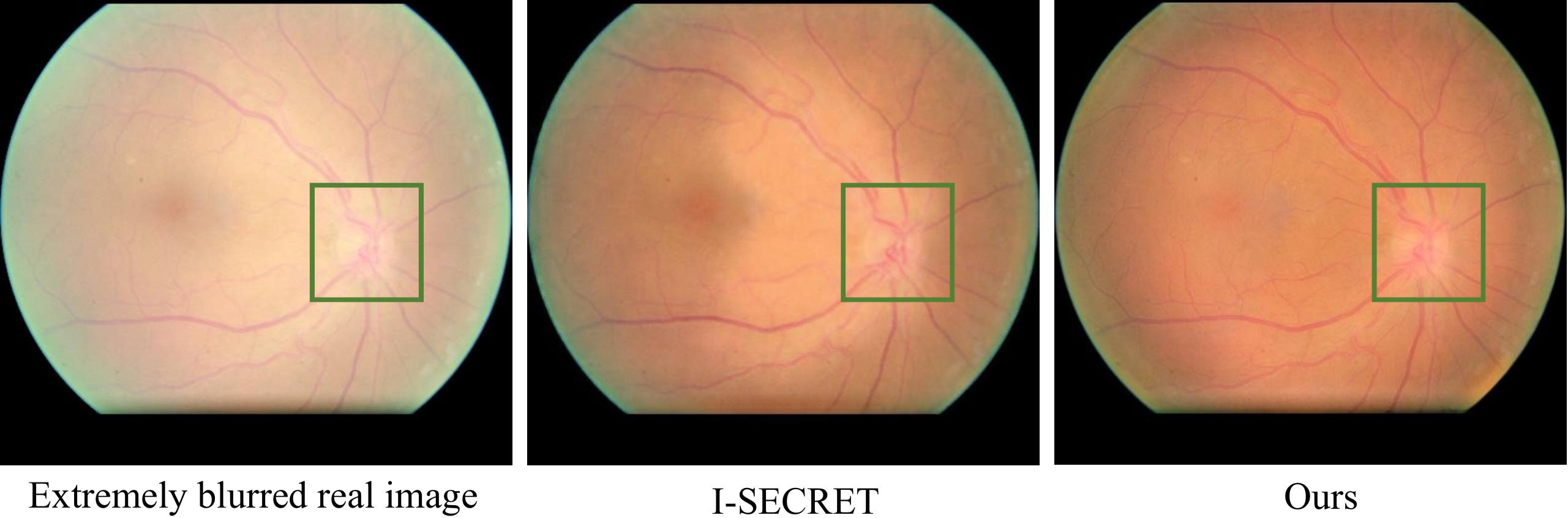}}
\caption{Visualization of failure cases. From the left to the right are an extremely blurred real low-quality fundus image, an enhanced image by I-SECRET~\cite{cheng2021secret}, and an enhanced image by our method. Neither I-SECRET~\cite{cheng2021secret} nor our method produces clear disc/cup and vessel details.} 
\label{fig6}
\end{figure}

Although our proposed method outperforms other strong competitors, our method may not well reconstruct the images with too much noise. For example, if the image is extremely over/under-exposured, the proposed method will not work. Also, the vessel and disc/cup can be preserved only with moderate level of noise. A visual example about this limitation is shown in Fig.~\ref{fig6}, which includes one extremely blurred real low-quality fundus image and the enhancement results from I-SECRET~\cite{cheng2021secret} and our method. Neither I-SECRET~\cite{cheng2021secret} nor our method produces clear disc/cup and vessel details. Fundus image enhancement under extreme noise condition is still a challenging problem, and will be investigated in our future work.

\section{Conclusion}
In this paper, we propose a new transferred MAGE-Net method by integrating synthetic and real-world low-quality fundus images for multi-stage fundus image enhancement guided by multi-attentions. Furthermore, we design an RSP module to preserve the anatomical retinal structures and integrate it with our mean teacher based multi-stage enhancement framework seamlessly.  Comprehensive experimental results demonstrate that our proposed method can simultaneously perform fundus image enhancement and reduce the domain gap between the synthetic and the authentic images. In addition, our method can boost the downstream tasks and assist clinical diagnosis for ophthalmologists and automated image analysis systems.

\appendices

\bibliographystyle{ieeetr}
\bibliography{mybibliography}
\end{document}